\documentclass[aps,prl,twocolumn,superscriptaddress,longbibliography]{revtex4-2}
\usepackage{graphicx}
\usepackage{amssymb,amsmath}
\usepackage{bm}
\usepackage{dcolumn}
\usepackage{float}
\usepackage[OT1]{fontenc} 
\usepackage{url}
\usepackage{mathrsfs}
\usepackage{slashed,comment}
\usepackage{color}
\usepackage{verbatim}
\usepackage{enumitem}
\usepackage{soul,physics,tikz}
\usepackage[driverfallback=dvipdfm]{hyperref}
\hypersetup{pdfpagemode=FullScreen,colorlinks=true,breaklinks,urlcolor=blue,linkcolor=blue,citecolor=blue}

\usepackage{subfigure}
\usepackage{amssymb}
\usepackage{bm}
\usepackage{graphicx}
\usepackage{amsmath,amssymb}

\usepackage{pifont}

\usetikzlibrary{positioning,calc,arrows.meta,shapes.geometric}

\begin{document}
 
  \title{Post-Selection Probability and Fidelity of Bidirectional Teleportation}

  \author{Ning Sun}
  \affiliation{State Key Laboratory of Surface Physics \& Department of Physics, Fudan University, Shanghai, 200438, China}

  \author{Lei Feng}
  \thanks{leifeng@fudan.edu.cn}
  \affiliation{State Key Laboratory of Surface Physics \& Department of Physics, Fudan University, Shanghai, 200438, China}
  \affiliation{Institute for Nanoelectronic devices and Quantum computing, Fudan University, Shanghai, 200438, China}
  \affiliation{Shanghai Key Laboratory of Metasurfaces for Light Manipulation, Shanghai, 200433, China}
  \affiliation{Hefei National Laboratory, Hefei 230088, China}

  \author{Pengfei Zhang}
  \thanks{PengfeiZhang.physics@gmail.com}
  \affiliation{State Key Laboratory of Surface Physics \& Department of Physics, Fudan University, Shanghai, 200438, China}
  \affiliation{Hefei National Laboratory, Hefei 230088, China}
  \date{\today}

  \begin{abstract}
  Understanding the scrambling of quantum information is central to many areas of quantum physics, including quantum thermalization, entanglement growth, and quantum information processing. Insights from these studies have, in turn, inspired the development of novel quantum protocols and algorithms. Recently, a bidirectional teleportation protocol was proposed to implement a digital SWAP operation between qubits by leveraging chaotic Hamiltonian evolution combined with measurement and post-selection. In this work, we provide a comprehensive study of two central quantities that characterize the protocol, the post-selection probability and the fidelity, taking into account possible errors in time-reversed dynamics. We show that these quantities can be expressed in terms of standard diagnostics in quantum dynamics, including the Loschmidt echo and its subsystem variant. The results unveil (1) the initial-state dependence of the fidelity and (2) the stability of the post-selection probability in integrable models. Our findings offer practical guidance for the implementation of the protocol on realistic quantum devices.

  \end{abstract}
    
  \maketitle

  \begin{figure}[t]
    \centering
    \includegraphics[width=0.98\linewidth]{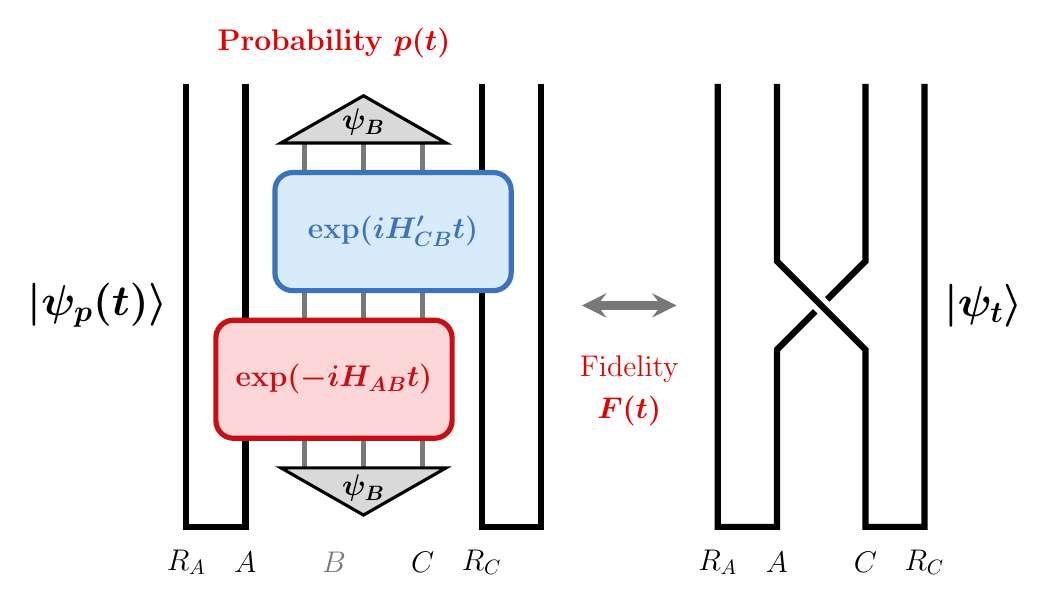}
    \caption{The schematic illustration of the bidirectional teleportation protocol between systems $A$ and $C$, with reference qubits $R_A$ and $R_C$. Here, we allow for errors between the forward and backward evolution Hamiltonians, denoted as $H$ and $H'$, respectively. It has been established that, under scrambled evolution with perfectly reversed dynamics, the protocol provides a probabilistic realization of the SWAP operation between $A$ and $C$ systems. The two central quantities governing the protocol are the post-selection probability $p(t)$ and the fidelity of the resulting quantum state $F(t)$. }
    \label{fig:schematics}
  \end{figure}

  \emph{ \color{blue}Introduction.---} In recent years, there has been growing interest in understanding quantum dynamics in many-body systems, as it provides a valuable framework for exploring fundamental questions such as the emergence of quantum statistical physics~\cite{PhysRevE.50.888,PhysRevA.43.2046} and the growth of complexity~\cite{Nielsen:2006cea,Susskind:2014rva,Brown:2017jil,PhysRevX.9.041017}. A central concept in this context is information scrambling, which describes the spread of initially localized quantum information throughout a system~\cite{Hayden:2007cs,Sekino:2008he,Shenker:2014cwa,Roberts:2014isa,Hosur:2015ylk}. It underlies a wide range of phenomena in condensed matter physics, high-energy physics, and quantum information theory. For example, the rate of information scrambling can distinguish thermalized systems, integrable systems, and many-body localized systems~\cite{2017SciBu..62..707F,2017AnP...52900318H,Swingle:2016jdj,PhysRevB.95.054201,2016arXiv160802765C,PhysRevB.95.165136}. Information scrambling also plays a pivotal role in the Hayden-Preskill thought experiment~\cite{Hayden:2007cs}, which examines how information thrown into an old black hole can be retrieved. In addition, experimental studies have been conducted on various quantum platforms~\cite{Li:2016xhw,CappellaroEmergentPrethermalizationSignatures2019,LiExperimentalObservationEquilibrium2020,2019arXiv190206628S,2021PhRvA.104a2402D,2022PhRvA.105e2232S,DuEmergentUniversalQuench2024a,Garttner:2016mqj,2019Natur.567...61L,ReyUnifyingScramblingThermalization2019b,LinkeExperimentalMeasurementOutofTimeOrdered2022,ChenInformationScramblingQuantum2021,DuanInformationScramblingDynamics2022,OliverProbingQuantumInformation2022,ZhaoProbingOperatorSpreading2022,ZobristConstructiveInterferenceEdge2025,VuleticTimereversalbasedQuantumMetrology2022,VuleticImprovingMetrologyQuantum2023,ThomasEnergyResolvedInformationScrambling2021,LiObservationQuantumInformation2024a,YouObservationAnomalousInformation2024a,Li:2025civ}, offering insights into complex quantum dynamics. 

  On the other hand, the rapid advancement of quantum science and technology poses new challenges for the design of novel quantum algorithms. Theoretical insights from information scrambling offer a promising avenue for addressing these challenges by enabling controlled manipulation of quantum information flow. A celebrated example is the Yoshida-Kitaev protocol for the Hayden-Preskill problem~\cite{Yoshida:2017non,PhysRevX.9.011006,Landsman:2018jpm,PhysRevResearch.2.043024}, which shares a similar spirit with wormhole teleportation~\cite{Gao:2019nyj,Brown:2019hmk,Nezami:2021yaq,PhysRevX.12.031013,Jafferis:2022crx,Zhou:2024osg,Liu:2024nhs}, a protocol motivated by the study of traversable wormholes~\cite{Gao:2016bin,Maldacena:2017axo,Maldacena:2018lmt}. In both protocols, a quantum state is teleported from one partner $A$ to another partner $C$ using the scrambling dynamics of many-body Hamiltonians, without requiring universal local control. Later, Vikram et al. extend this idea by introducing a novel bidirectional teleportation protocol that probabilistically realizes a SWAP operation between $A$ and $C$ using scrambling dynamics~\cite{Vikram:2026wdg}. The protocol is validated through explicit numerical simulations.
  
  In this work, we analyze the performance of the protocol by focusing on two central quantities: the probability of post-selection, denoted $p(t)$, and the fidelity of the post-selected quantum state, denoted $F(t)$. Our analysis accounts for errors in the backward-evolution Hamiltonians, which are inevitable in realistic experiments~\cite{Li:2025civ,Liu:2026nnw}. We show that $p(t)$ and $F(t)$ can be expressed in terms of the Loschmidt echo~\cite{PhysRevA.30.1610,Goussev:2012swt} and its subsystem variant~\cite{Chen:2020wiq,Karch:2025wli}, which probe entropy growth and quantum chaos in many-body systems. These relations have direct implications for the initial-state dependence of the fidelity and the stability of the post-selection probability. Consequently, our results are of fundamental interest for the experimental realization of the bidirectional teleportation protocol and, more generally, for understanding the ability of quantum information processing in quantum many-body dynamics.

  \emph{ \color{blue}Protocol.---} We consider the scenario where each partner, $A$ and $C$, consists of $N$ qubits. The goal is to implement bidirectional teleportation, or a SWAP operation, between their quantum states. However, digital quantum operations on $A$ and $C$ are unavailable, and the only allowed dynamics involve coupling them to a large system $B$ via many-body Hamiltonians. The number of qubits in system $B$ is denoted by $M$, with $M \gg N$. In our analysis, rather than specifying particular initial states for $A$ and $C$~\cite{Vikram:2026wdg}, we entangle them with reference qubits, $R_A$ and $R_C$. The initial state of the total system reads
  \begin{equation}\label{eqn:ini}
  |\psi_0\rangle=|\text{EPR}\rangle_{AR_A}\otimes|\psi_B\rangle_B \otimes|\text{EPR}\rangle_{CR_C}.
  \end{equation}
  Here, $|\text{EPR}\rangle_{AR_A}=2^{-N/2}\sum_{i=1}^{2^N}|i\rangle_A\otimes |i\rangle_{R_A}$ denotes the Einstein-Podolsky-Rosen (EPR) state in the computational basis. The state $|\psi_B\rangle$ represents a general initial pure state of the system $B$. 

  The protocol is illustrated in FIG.~\ref{fig:schematics}. First, $A$ and $B$ are coupled, and the system evolves to time $t$ under the Hamiltonian $H_{AB}$. Subsystem $A$ is then decoupled from $B$, and $C$ is coupled to the system $B$ instead. Ideally, $BC$ is subsequently evolved backward under the Hamiltonian $H_{CB}$ for a time $t$, which has exactly the same form as $H_{AB}$. However, errors in the time-reversed dynamics are inevitable in realistic experiments. Therefore, we allow errors in the backward evolution Hamiltonian and denote it as $H'_{CB}$. The resulting quantum state, parametrized by the evolution time $t$, reads
  \begin{equation}
  |\psi(t)\rangle=e^{iH_{CB}'t}e^{-iH_{AB}t}|\psi_0\rangle.
  \end{equation}
  Finally, the subsystem $B$ is measured using the projection operators $\{P_B,, 1-P_B\}$, where $P_B = |\psi_B\rangle_B \,_B\langle \psi_B|$. We post-select the states in which subsystem $B$ returns to $|\psi_B\rangle_B$, thereby defining an (unnormalized) pure quantum state for the remaining system as
  \begin{equation}
  |\psi_{\text{p}}(t)\rangle=\,_B\langle \psi_B|e^{iH_{CB}'t}e^{-iH_{AB}t}|\psi_0\rangle
  \end{equation}
  The probability for the post-selection is given by $p(t)=\langle \psi_{\text{p}}(t)|\psi_{\text{p}}(t)\rangle$. This resulting state is compared to the target quantum state, defined by applying the SWAP operation between $A$ and $C$ on the initial state \eqref{eqn:ini}, with subsystem $B$ neglected:
  \begin{equation}
  |\psi_\text{t}\rangle=|\text{EPR}\rangle_{CR_A}\otimes|\text{EPR}\rangle_{AR_C}.
  \end{equation}
  We define the fidelity of the protocol as the overlap between the post-selected state and the target state: $F(t)=|\langle \psi_{\text{t}}|\psi_{\text{p}}(t)\rangle|^2/p(t)$. Here, $p(t)$ accounts for the normalization. In Ref.~\cite{Vikram:2026wdg}, it was established that in the errorless limit, approximating the evolution operator as a Haar-random unitary yields $p_{\text{Haar}} \approx 2^{-2N}$ and $F_{\text{Haar}} \approx 1$, demonstrating the validity of the protocol in the ideal case. 

  \begin{figure}[t]
    \centering
    \includegraphics[width=0.75\linewidth]{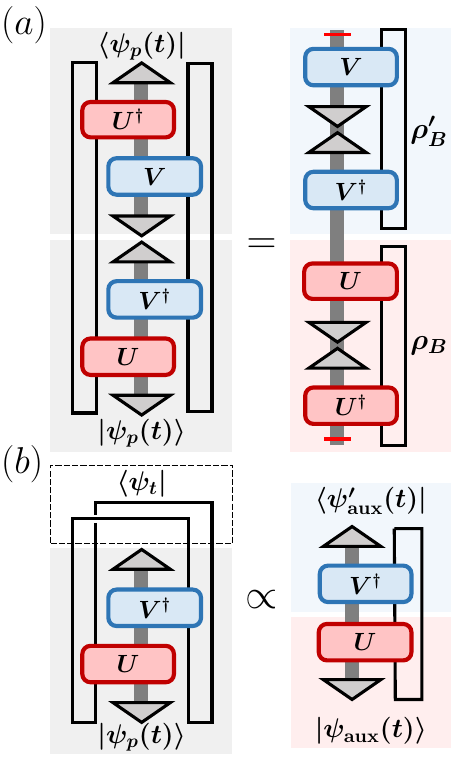}
    \caption{The diagrammatic proof of the main relations: (1) between the probability and the subsystem echo, $p(t) = L_B(t)$, and (2) between the fidelity and the Loschmidt echo, $F(t)p(t) = 2^{-2N} L(t)$. Here, the overall factor arises from the normalization of the EPR state, which is omitted in the diagrammatic representation. For conciseness, we introduce unitary evolutions $U = e^{-i H_{AB} t}$ and $V = e^{-i H'_{AB} t}$, and represent the qubits in $B$ by a single thick solid line. The red slash indicates periodic boundary conditions. }
    \label{fig:proof}
  \end{figure}

  \emph{ \color{blue}General Relations.---} However, Hamiltonian evolutions are far from Haar random due to energy conservation~\cite{Roberts:2016hpo}, and the effect of errors can become significant. Motivated by this, we derive a general relation between $p(t)$, $F(t)$, and Loschmidt echoes in an auxiliary quantum dynamical process. The diagrammatic derivation is shown in FIG.~2, following the standard notation in quantum information science~\cite{nielsen2000quantum}. Here, we only describe the resulting relation. The auxiliary quantum dynamics involves only systems $A$, $R_A$ and $B$. The system is initialized in
  \begin{equation}
  |\psi_{0,\text{aux}}\rangle=|\text{EPR}\rangle_{AR_A}\otimes|\psi_B\rangle_B.
  \end{equation}
  Next, we evolve the state to time $t$ under either Hamiltonian $H_{AB}$ or $H_{AB}'$. The resulting quantum states are denoted by $|\psi_{\text{aux}}(t)\rangle=e^{-iH_{AB}t}|\psi_{0,\text{aux}}\rangle$ and $|\psi_{\text{aux}}'(t)\rangle=e^{-iH_{AB}'t}|\psi_{0,\text{aux}}\rangle$, respectively. The deviation between two quantum states can be detected using the Loschmidt echo~\cite{PhysRevA.30.1610,Goussev:2012swt} (see FIG.~\ref{fig:proof} (b) for a diagrammatic representation):
  \begin{equation}
  \begin{aligned}
  L(t)&=|\langle \psi_{\text{aux}}'(t)|\psi_{\text{aux}}(t)\rangle|^2\\&=|\langle \psi_{0,\text{aux}}'|e^{iH_{AB}'t}e^{-iH_{AB}t}|\psi_{0,\text{aux}}\rangle|^2,
  \end{aligned}
  \end{equation}
  which probes how small errors in the Hamiltonian perturb the entire quantum state, thereby characterizing quantum chaotic behavior. In the errorless limit $H_{AB}=H'_{AB}$, we have $L(t)=1$ for arbitrary time $t$. We further introduce a subsystem variant of the Loschmidt echo, which has been considered in discussions of information retrieval~\cite{Chen:2020wiq}. We first trace out the subsystem $AR_A$ to obtain the reduced density matrix for system $B$ as $\rho_B(t)=\text{tr}_{AR_A}[|\psi_{\text{aux}}(t)\rangle \langle \psi_{\text{aux}}(t)|]$, and similarly $\rho_B'(t)=\text{tr}_{AR_A}[|\psi_{\text{aux}}'(t)\rangle \langle \psi_{\text{aux}}'(t)|]$. We then define the subsystem Loschmidt echo by evaluating the density matrix overlap (see FIG.~\ref{fig:proof} (a) for a diagrammatic representation):
  \begin{equation}
  L_B(t)=\text{tr}_B[\rho_B'(t)\rho_{B}(t)].
  \end{equation}
  In the errorless limit, this reduces to the purity of subsystem $B$, which is bounded by $L_B(t)\geq 2^{-2N}$, since the total system is in a pure state.

  We derive the relation between the post-selection probability $p(t)$ and the fidelity $F(t)$ in the bidirectional teleportation protocol, and the Loschmidt echo $L(t)$ and its subsystem variant $L_B(t)$ in the auxiliary quantum dynamics, by representing each quantity using its corresponding diagrammatic representation and identifying equivalent diagrams (see FIG.~\ref{fig:proof}). The result reads
  \begin{equation}\label{eq:general}
  p(t)=L_B(t),\ \ \ \ \ F(t)=2^{-2N}L(t)/L_B(t).
  \end{equation}
  These are the main results of the manuscript. They are consistent with the errorless Haar-random unitary protocol, which has maximal coherence $L(t)=1$ and minimum purity $L_B(t)\approx2^{-2N}$, leading to $p(t)\approx2^{-2N}$ and $F(t)\approx 1$. 

  \emph{ \color{blue}Analysis.---} We now consider more general errorless Hamiltonian dynamics. The results \eqref{eq:general} reduce to
  \begin{equation}\label{eq:resnoerr}
  p(t)=L_B(t),\ \ \ \ \ F(t)=2^{-2N}/L_B(t).\ \ \ \text{(errorless)}
  \end{equation}
  Let us assume that the Hamiltonian $H_{AB}$ is chaotic. In this case, we expect the system to exhibit quantum thermalization as $t$ becomes sufficiently long. Since $N \ll M$, the subsystem $A$ thermalizes to a thermal density matrix, which predicts that $L_B(t)=2^{-N}e^{-N s_{\text{th}}(\beta)}$, where $s_{\text{th}}(\beta)$ is the thermal R\'enyi entropy per qubit at an effective inverse temperature $\beta$ determined through the energy
  \begin{equation}\label{eq:ETH}
  \langle \psi_{0,\text{aux}}|H_{AB}|\psi_{0,\text{aux}}\rangle=\frac{\text{tr}[e^{-\beta H_{AB}}H_{AB}]}{\text{tr}[e^{-\beta H_{AB}}]}.
  \end{equation}
  Therefore, achieving maximal fidelity $F(t)=1$ requires maximal thermal entropy $s_{\text{th}}(\beta)=\ln 2$, corresponding to $\beta=0$. Consequently, the initial state $|\psi_B\rangle$ should be chosen such that $\langle \psi_{0,\text{aux}}|H_{AB}|\psi_{0,\text{aux}}\rangle=\text{tr}[H_{AB}]/2^{2N+M}$. This condition ensures that subsystem $B$ begins in an infinite-temperature state and can fully scramble information throughout the entire Hilbert space.

  \begin{figure}[t]
    \centering
    \includegraphics[width=0.98\linewidth]{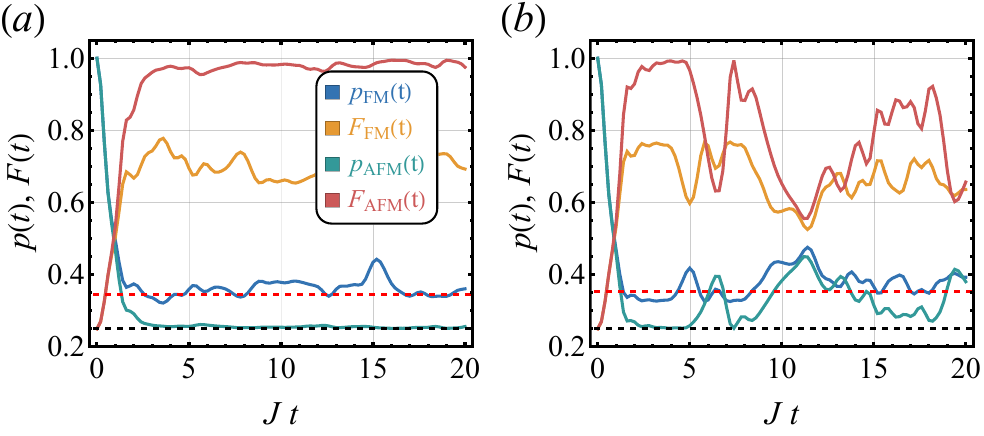}
    \caption{We present the numerical results for the quantum Ising model in the absence of errors in the backward evolution. We fix $N = 1$, $M = 8$, and $g/J = 1.05$, and consider (a) $h = 0.5$ (chaotic~\cite{Hosur:2015ylk,Li:2016xhw}) and (b) $h = 0$ (integrable). We present the post-selection probability and the fidelity as functions of time $t$ for both ferromagnetic and antiferromagnetic initial states, denoted as FM and AFM, respectively. The dashed lines represent the theoretical prediction $p(t) =  {e^{-s_{\text{th}}}}/{2}$, as described in the main text. }
    \label{fig:num1}
  \end{figure}

  Next, we analyze the effect of errors. We expect the Loschmidt echo $L(t)$ to decay over time $t$ rapidly due to the amplification of small errors by quantum chaos~\cite{Li:2025civ}. Similarly, the subsystem echo $L_B(t)$ deviates from the purity and is no longer lower-bounded by $2^{-2N}$. For chaotic systems, the long-time limit can be modeled by assuming that the evolution operators $e^{-iH_{AB}t}$ and $e^{-iH_{AB}'t}$ become independent samples of the Haar-random unitary. Performing the average over the Haar unitary ensemble~\cite{Roberts:2016hpo} leads to $L_B(t)=2^{-M}$ and $L(t)=2^{-2N-M}$, which gives
  \begin{equation}\label{eq:approx_long}
  p(\infty)=2^{-M},\ \ \ \ \ F(\infty)=2^{-4N}.
  \end{equation}
  Since the post-selection probability is exponentially small for a large system $B$, the protocol is unlikely to succeed even for a small number of qubits to be teleported. 

  The above analysis further motivates us to consider integrable Hamiltonians as an alternative candidate for bidirectional teleportation with errors in the backward evolution. The reasoning is as follows: (1) The errorless scenario only requires maximal entanglement between the small subsystem $AR_A$ and the remaining system $B$, which can also be achieved without quantum chaos. For example, if the system $AB$ is evolved under a brick-wall quantum circuit of SWAP gates, the purity reaches its minimum immediately after a single layer of the circuit \cite{Bertini:2019gbu,Bertini:2019wkb,PhysRevLett.123.210601,Akhtar:2024nwc,Song:2024ukm}, even though SWAP gates merely relabel qubits and are therefore non-chaotic. (2) Integrability generally suppresses the amplification of errors in time-reversed dynamics, leading to a larger post-selection probability compared to chaotic Hamiltonians. Therefore, we expect the protocol to become more stable when integrable Hamiltonians are employed in practical systems.

  \begin{figure}[t]
    \centering
    \includegraphics[width=0.75\linewidth]{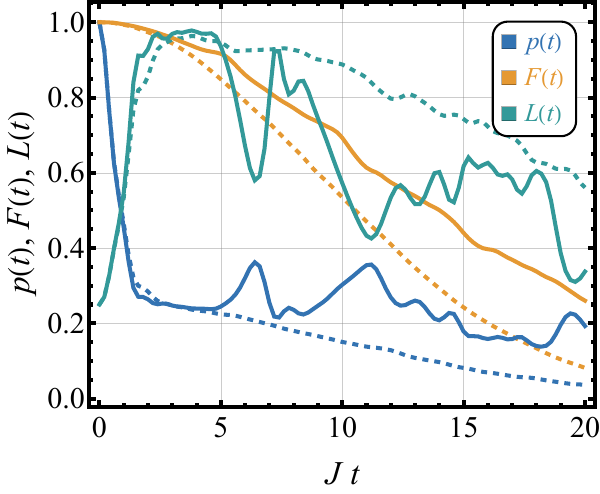}
    \caption{We present the numerical results for the quantum Ising model with an error $\epsilon = 0.05$. We fix $N = 1$, $M = 8$, and $g/J = 1.05$, and consider the antiferromagnetic initial state. The solid lines correspond to the integrable case with $h = 0$, while the dashed lines represent the chaotic case with $h = 0.5$. }
    \label{fig:num2}
  \end{figure}

  \emph{ \color{blue}Numerics.---} We provide numerical justification for the above analysis by focusing on the quantum Ising model, with Hamiltonian
  \begin{equation}
  H_{AB}=J \sum_{i=1}^{L-1} Z_i Z_{i+1}+g\sum_{i=1}^LX_i+h\sum_{i=1}^LZ_i.
  \end{equation}
  The system contains $L = N + M$ sites with open boundary conditions, where the first $N$ sites represent system $A$ and the remaining sites represent system $B$. We set $J = 1$ as the energy unit. The system is integrable for either $g = 0$ or $h = 0$, and is otherwise chaotic. Information scrambling in this model has been explored experimentally in~\cite{Li:2016xhw}, revealing a near-monotonic decay of the out-of-time-order correlator in the chaotic case and persistent oscillations in the integrable case.

  We first focus on the errorless limit and study the effects of different initial states. We compute the subsystem Loschmidt echo $L_B(t)$ for a given initial state $|\psi_B\rangle$, obtaining both the probability $p(t)$ and the fidelity $F(t)$. We consider two different initial states: (1) a ferromagnetic state along the $x$ direction $|\psi_B^{(1)}\rangle=\ket{+++\cdots}$ and (2) a anti-ferromagnetic state $|\psi_B^{(2)}\rangle=\ket{+-+\cdots}$. The Hamiltonian is traceless, and the infinite-temperature state has vanishing energy. The ferromagnetic state yields an initial state $|\psi_{0,\text{aux}}\rangle$ with extensive energy $E = gM$, and thus corresponds to a finite-temperature state. By contrast, for even $M$, the antiferromagnetic state has zero energy and corresponds to an infinite effective temperature. Based on our previous analysis, we expect the anti-ferromagnetic state to exhibit better performance than the ferromagnetic state. 

  The numerical results without errors in the backward evolution are presented in FIG.~\ref{fig:num1} for $N=1$ and $M=8$. The integrable case exhibits persistent oscillations that are significantly more pronounced than in the chaotic case. Nevertheless, for both chaotic and integrable dynamics, we find that the post-selection probability can reach $p(t)\approx 1/4$ for the antiferromagnetic initial state, corresponding to $F(t)\approx 1$. In contrast, the ferromagnetic initial state yields a higher probability, which, according to \eqref{eq:resnoerr}, leads to a lower fidelity. We further compare the numerical results with the theoretical prediction $p(t)=e^{-s_{\text{th}(\beta)}}/2$, shown as dashed lines. Here, $\beta$ is first determined numerically through the relation \eqref{eq:ETH}, and $s_{\text{th}}$ is obtained by computing the thermal R\'enyi entropy $S_{\text{th}}=-\ln\text{tr}[\rho_{\text{th}}^2]=(M+N)s_{\text{th}}$ for the corresponding thermal density matrix $\rho_{\text{th}}=e^{-\beta H_{AB}}/{\text{tr}[e^{-\beta H_{AB}}]}$. For the chaotic case, the predictions are in excellent agreement with the numerical data.

  We then turn to the scenario with errors in the backward evolution. As a simple example, we consider the case where $\hat{H}'_{AB}$ is still described by the quantum Ising model, but with a modified transverse field $g' = g + \epsilon$. The numerical results for anti-ferromagnetic intital states are presented in FIG.~\ref{fig:num2}. When an error is present, the Loschmidt echo decays over time, and the subsystem echo is no longer bounded by $2^{-2N}$. For integrable systems, the decay of the (subsystem) Loschmidt echo is significantly slower, resulting in a larger post-selection probability. For example, an enhancement by a factor of 2.6 is observed at $Jt \approx 18$. At the same time, both the integrable system can reach a similar fidelity, when choosing evolution times near the peak of the oscillations. This clearly demonstrate the usefulness of integrable systems in practical systems. 

  \emph{ \color{blue}Discussions.---} In this work, we analyze the bidirectional teleportation protocol by connecting the post-selection probability and fidelity to standard quantum dynamics diagnostics: the subsystem and full Loschmidt echoes (Eq.~\eqref{eq:general}). These relations hold for arbitrary Hamiltonians and backward-evolution errors, providing a unified framework for assessing performance. In the errorless limit, achieving optimal fidelity requires the initial state to have infinite effective temperature, minimizing subsystem purity. When time-reversal errors are present, the post-selection probability decays with the Loschmidt echo, with chaotic systems amplifying errors much faster than integrable ones. Remarkably, integrable Hamiltonians can still produce sufficient entanglement for near-unit fidelity while their stability under imperfect time reversal significantly enhances post-selection probability, making them a promising candidate for near-term quantum devices.

  Several interesting directions emerge from this work. First, experimental realizations on platforms such as trapped ions or superconducting qubits could test the performance of the protocol, particularly highlighting the stability advantages of integrable Hamiltonians in the presence of experimental noise. Second, it is interesting to investigate whether errors can be mitigated in certain classes of quantum systems, similar to recent approaches for measuring the out-of-time-order correlator~\cite{Li:2025civ}, potentially employing theoretical tools such as scramblon theory~\cite{Gu:2021xaj,Stanford:2021bhl}. Finally, one may explore a broader design principle for scrambling-based quantum protocols: instead of relying exclusively on chaotic dynamics, the tunability between integrable and chaotic regimes could be exploited to balance fidelity and robustness, an idea that could be applied in other settings, such as wormhole teleportation.

  \vspace{5pt}
  \textit{Acknowledgement.}
  This project is supported by the NSFC under grant 12374477 (PZ), the Shanghai Rising-Star Program under grant number 24QA2700300 (PZ), the Shanghai Municipal Science and Technology Major Project Grant No. 24DP2600100 (NS and LF), Co-research Program under Grant No. 25LZ2601000 (NS and LF), the Quantum Science and Technology--National Science and Technology Major Project under Grant No. 2023ZD0300900 (LF), 2024ZD0300101 (PZ), 2025ZD0300100 (NS and LF) and 2025ZD0300101 (NS and LF). PZ is also supported by the Xuemin Institute of Advanced Studies at Fudan University.

\bibliography{ref.bib}

\end{document}